\documentclass[english]{revtex4}
\usepackage[T1]{fontenc}
\usepackage[latin9]{inputenc}
\usepackage{float}
\usepackage{relsize}
\usepackage{graphicx}
\usepackage{esint}

\makeatletter

\DeclareRobustCommand{\greektext}{%
  \fontencoding{LGR}\selectfont\def\encodingdefault{LGR}}
\DeclareRobustCommand{\textgreek}[1]{\leavevmode{\greektext #1}}
\DeclareFontEncoding{LGR}{}{}

\DeclareRobustCommand{\lyxmathsym}[1]{\ifmmode\begingroup\def\b@ld{bold}
  \def\rmorbf##1{\ifx\math@version\b@ld\textbf{##1}\else\textrm{##1}\fi}
  \mathchoice{\hbox{\rmorbf{#1}}}{\hbox{\rmorbf{#1}}}
  {\hbox{\smaller[2]\rmorbf{#1}}}{\hbox{\smaller[3]\rmorbf{#1}}}
  \endgroup\else#1\fi}

\makeatother

\usepackage{babel}

\begin{document}

\title{Entropy for Black Holes in the Deformed Horava-Lifshitz Gravity}

\author{Andres Castillo}

\affiliation{Universidad Nacional de Colombia. Departamento de Física.}

\author{Alexis Larrañaga}

\affiliation{Universidad Nacional de Colombia. Observatorio Astronómico Nacional
(OAN)}

\affiliation{Universidad Distrital Francisco Jose de Caldas. Facultad de Ingeniería.}
\begin{abstract}
We study the entropy of black holes in the deformed Horava-Lifshitz
gravity with coupling constant $\lyxmathsym{\textgreek{l}}$. For
$\lyxmathsym{\textgreek{l}}=1$, the black hole resembles the Reissner-Norstrom
black hole with a geometric parameter acting like the electric charge.
Therefore, we obtain some differences in the entropy when comparing
with the Schwarzschild black hole. Finally, we study the heat capacity
and the thermodynamical stability of this solution.
\end{abstract}
\maketitle


\section{Introduction}

Recently, Horava \cite{horava} proposed a non-relativistic renormalisable
theory of gravity that reduces to Einstein\textquoteright{}s general
relativity at large scales. This theory is named Horava-Lifshitz theory
and has been studied in the literature for its applications to cosmology
\cite{cosmology} and black holes \cite{blackholes}. However, this
proposal introduces back a non-equality of space and time. Therefore,
in this approach, space and time exhibit Lifshitz scale invariance
$t\rightarrow l^{z}t$ and $x^{i}\rightarrow lx^{i}$ with $z\geq1$.
Moreover, the theory is not invariant under the full diffeomorphism
group of General Relativity (GR), but rather under a subgroup of it.
This fact is manifest using the standard ADM splitting. 

However, the Horava-Lifshitz theory goes to standard GR if the coupling
$\lyxmathsym{\textgreek{l}}$ that controls the contribution of the
trace of the extrinsic curvature has the specific value $\lambda=1$.
For generic values of \textgreek{l}, the theory does not exhibits
the full 4D diffeomorphism invariance at large distances and it is
possible to obtain deviations from GR. Therefore, it is interesting
to confront this type of non-relativistic theory with experimental
and observational data.

Using the (3 + 1)-dimensional ADM formalism, the general metric can
be written as 

\begin{equation}
ds^{2}=-N^{2}dt^{2}+g_{ij}(dx^{i}+N^{i}dt)(dx^{j}+N^{j}dt)\label{eq:genmetric}\end{equation}

where $g_{ij}$, $N$ and $N^{i}$ are the dynamical fields of scaling
mass dimensions 0, 0, 2, respectively. The Einstein-Hilbert action
can be expressed as\begin{equation}
S_{EH}=\frac{1}{16\pi G}\int d^{4}x\sqrt{g}N\lbrace\left(K_{ij}K^{ij}-K^{2}\right)+R-2\Lambda\rbrace,\label{eq:EHaction}\end{equation}
where $G$ is Newton\textquoteright{}s constant and the extrinsic
curvature $K_{ij}$ takes the form 

\begin{equation}
K_{ij}=\frac{1}{2N}\left(\dot{g}_{ij}-\nabla_{i}N_{j}-\nabla_{j}N_{i}\right)\end{equation}

with a dot denoting derivative with respect to $t$ and covariant
derivatives defined with respect to the spatial metric $g_{ij}$.
On the other hand, the action of Horava-Lifshitz theory is given by
\cite{horava}

\begin{eqnarray}
S_{HL} & = & \int dtd^{3}x\sqrt{g}N\left[\frac{2}{\kappa^{2}}\left(K_{ij}K^{ij}-\lambda K^{2}\right)+\frac{\kappa^{2}\mu^{2}(\tilde{\Lambda}R-3\tilde{\Lambda}^{2})}{8(1-3\lambda)}+\frac{\kappa^{2}\mu^{2}(1-4\lambda)}{32(1-3\lambda)}R^{2}\right.\nonumber \\
 &  & \left.-\frac{\kappa^{2}\mu^{2}}{8}R_{ij}R^{ij}+\frac{\kappa^{2}\mu}{2\omega^{2}}\epsilon^{ijk}R_{il}\nabla_{j}R_{k}^{l}-\frac{\kappa^{2}}{2\omega^{4}}C_{ij}C^{ij}\right],\label{eq:LHaction}\end{eqnarray}

where $\kappa^{2}$, $\lyxmathsym{\textgreek{l}}$, $\lyxmathsym{\textgreek{w}}$
are dimensionless constant parameters while $\mu$ and $\lyxmathsym{\textgreek{L}}$
are constant parameters with mass dimensions $\left[\mu\right]=1$,
$\left[\Lambda\right]=2$. The object $C_{ij}$ is called the Cotten
tensor, defined by

\begin{equation}
C^{ij}=\epsilon^{ijk}\nabla_{k}R_{l}^{j}-\frac{1}{4}\epsilon^{ijk}\partial_{k}R.\end{equation}

Comparing the action to that of general relativity, one can see that
the speed of light, Newton\textquoteright{}s constant and the cosmological
constant are

\begin{eqnarray}
c & = & \frac{\kappa^{2}\mu}{4}\sqrt{\frac{\tilde{\Lambda}}{1-3\lambda}}\\
G & = & \frac{\kappa^{2}c}{32\pi}\\
\Lambda & = & \frac{3}{2}\tilde{\Lambda}.\end{eqnarray}

Note that if $\lyxmathsym{\textgreek{l}}=1$, the first two terms
in (\ref{eq:LHaction}) could be reduced to the Einstein\textquoteright{}s
general relativity action (\ref{eq:EHaction}). However, in Horava-Lifshitz
theory, $\lyxmathsym{\textgreek{l}}$ is a dynamical coupling constant,
susceptible to quantum correction.

The static, spherically symmetric solutions have been found in \cite{blackholes}.
Because of the prescence of a cosmological constant, solutions for
\textgreek{l} = 1 are asymptotically AdS and have some interest because
the AdS/CFT correspondence. These solutions have also been extended
to general topological black holes\cite{topoBH}, in which the 2-sphere
that acts as horizon has been generalized to two dimensional constant
curvature spaces.

\section{Horava-Lifshitz Black Hole}

Now we will introduce the black hole solution in the limit of $\tilde{\Lambda}\rightarrow0$
and its thermodynamic properties. Considering $N_{i}=0$ (spherically
symmetric solutions) in (\ref{eq:genmetric}) we obtain the metric
ansatz 

\begin{equation}
ds^{2}=-N^{2}(r)dt^{2}+\frac{dr^{2}}{f(r)}+r^{2}d^{2}\Omega.\end{equation}

Using this line element and after the angular integration, the Lifshitz-Horava
lagrangian reduces to 

\begin{equation}
\tilde{\mathcal{L}}_{1}=\frac{\kappa^{2}\mu^{2}N}{8(1-3\lambda)\sqrt{f}}\left(\frac{\lambda-1}{2}f'^{2}-\frac{2\lambda(f-1)}{r}f'+\frac{(2\lambda-1)(f-1)^{2}}{r^{2}}-2\omega(1-f-rf')\right),\end{equation}

where 

\begin{equation}
\omega=\frac{8\mu^{2}(3\lambda-1)}{\kappa^{2}}.\end{equation}

For $\lyxmathsym{\textgreek{l}}=1$ , we have $\lyxmathsym{\textgreek{w}}=\frac{16\mu^{2}}{\kappa^{2}}$
and the functions $f$ and $N$ can be determined as \cite{thermodeformedbh}

\begin{equation}
N^{2}=f(r)=1+\omega r^{2}-\sqrt{r(\omega^{2}r^{3}+4\omega M)},\end{equation}
where $M$ is an integration constant that will be related to the
mass of the black hole. The condition $f\left(r_{\pm}\right)=0$ defines
the radii of the horizons

\begin{equation}
r_{\pm}=M\left[1\pm\sqrt{1-\frac{1}{2\omega M^{2}}}\right].\end{equation}
This equation shows that

\begin{equation}
M^{2}\geq\frac{1}{2\omega}\end{equation}

in order to have a black hole. The equality corresponds to the extremal
black hole in which the degenerate horizon has a radius

\begin{equation}
r_{e}=M_{e}=\frac{1}{\sqrt{2\omega}}.\end{equation}
In order to compare with Schwarzschild's slution, we define a new
parameter $\alpha$ as

\begin{equation}
\alpha=\frac{1}{2\omega},\end{equation}
so the function $f$ becomes

\begin{equation}
f(r)=\frac{2r^{2}-4Mr+2\alpha}{r^{2}+2\alpha+\sqrt{r^{4}+8\alpha Mr}}.\label{eq:fofalpha}\end{equation}
Note that $f\rightarrow2\left(1-\frac{2M}{r}\right)$ as $\alpha\rightarrow0$,
i.e. that we recover Schwarzschild's solution when $\omega\rightarrow\infty$.
Now, the horizon radii become

\begin{equation}
r_{\pm}=M\pm\sqrt{M^{2}-\alpha},\end{equation}
showing an incredible resemblance with the Reissner-Nordstrom solution
in which the horizons are defined by $r_{\pm}=M\pm\sqrt{M^{2}-Q^{2}}$,
i.e. that the parameter $\alpha$ can be associated with the electric
charge. In terms of $\alpha$, the extremal black hole i characterized
by the degenerate horizon

\begin{equation}
r_{e}=M_{e}=\sqrt{\alpha}.\end{equation}

Since this spacetime is spherically symmetric, the temperature of
the black hole can be calculated as

\begin{equation}
T=\frac{\kappa}{2\pi}=\frac{1}{4\pi}\left.\frac{\partial f(r)}{\partial r}\right|_{r=r_{+}},\end{equation}
or using equation (\ref{eq:fofalpha}),

\begin{equation}
T=\frac{r_{+}^{2}-\alpha}{4\pi(r_{+}^{3}+2\alpha r_{+})}.\label{eq:temperature}\end{equation}
Note that this temperature becomes Schwarzschild's black hole temperature
$T_{s}=\frac{1}{4\pi r_{+}}$ for $\alpha=0$. Evenmore, in the extremal
case, $r_{+}=r_{e}=\sqrt{\alpha}$ the temperature vanishes.

\section{Entropy}

Using the condition $f\left(r_{+}\right)=0$, the mass function is
given by

\begin{equation}
M(r_{+},\omega)=\frac{1+2\omega r_{+}^{2}}{4\omega r_{+}},\end{equation}
or in terms of the parameter $\alpha$,

\begin{equation}
M(r_{+},\alpha)=\frac{\alpha+r_{+}^{2}}{2r_{+}}\label{eq:massformula}\end{equation}

In Einstein\textquoteright{}s general relativity, entropy of black
hole is always given by one quarter of black hole horizon area, but
in higher derivative gravities, in general, the area formula breaks
down. Therefore, we will obtain the black hole entropy by using the
first law of black hole thermodynamics, assuming that this black hole
is a thermodynamical system and the first law keeps valid,

\begin{equation}
dM=TdS.\end{equation}

Note that we do not associate a thermodynamical character to the parameter
$\alpha$. Integrating this relation yields 

\begin{equation}
S=\int\frac{dM}{T}+S_{0},\end{equation}
where $S_{0}$ is an integration constant, which should be fixed by
physical consideration. Since the mass of the black hole is a function
of $r_{+}$ we can write

\begin{equation}
S=\int\frac{1}{T}\frac{\partial M}{\partial r_{+}}dr_{+}+S_{0}.\end{equation}
Using the temperature (\ref{eq:temperature}) and the mass formula
(\ref{eq:massformula}), we obtain the entropy

\begin{equation}
S=S_{0}+\pi\left(r_{+}^{2}+4\alpha\ln(r_{+})\right),\end{equation}
that is similar to the entropy obtained for topological black holes
in \cite{topoBH}. Since we want that this entropy becomes one quarter
of black hole horizon area for Schwarzschild's limit (i.e. $\alpha=0$),
we conclude that the integration constant is $S_{0}=0$, giving the
entropy

\begin{equation}
S=\pi r_{+}^{2}+4\pi\alpha\ln(r_{+}).\label{eq:entropy}\end{equation}

\begin{figure}[H]
\begin{centering}
\includegraphics[scale=0.4]{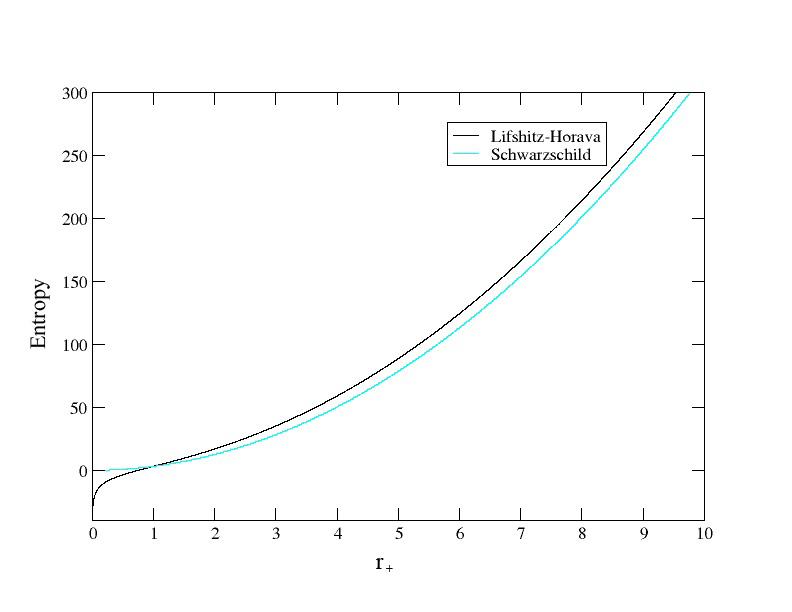}
\par\end{centering}

\caption{\label{fig:Entropy}Entropy as a function of the radius of outer horizon
with $\omega=1$ , i.e. $\alpha=\frac{1}{2}$. }

\end{figure}

In Figure \ref{fig:Entropy} it is shown the behavior of the entropy
as a function of the horizon radius in the case $\alpha=\frac{1}{2}$,
i.e. $\omega=1$ for the Horava-Lifshitz black hole as well as for
Schwarzschild's solution. Note how the curves coincide for large $r_{+}$
but there is a significant difference near $r_{+}=0$, because the
Horava-Lifshitz entropy diverges at this point. However, remember
that the Horava-Lifshitz black hole exists for values of $M\geq\sqrt{\alpha}=\frac{1}{\sqrt{2\omega}}$,
therefore, it can only have horizon radii satisfying $r_{+}\geq r_{e}=M_{e}$,
or $r_{+}\geq\frac{1}{\sqrt{2\omega}}$. Thus, the radius $r_{+}=0$
is not allowed and there is no entropy divergence.

\begin{figure}[H]
\begin{centering}
\includegraphics[scale=0.4]{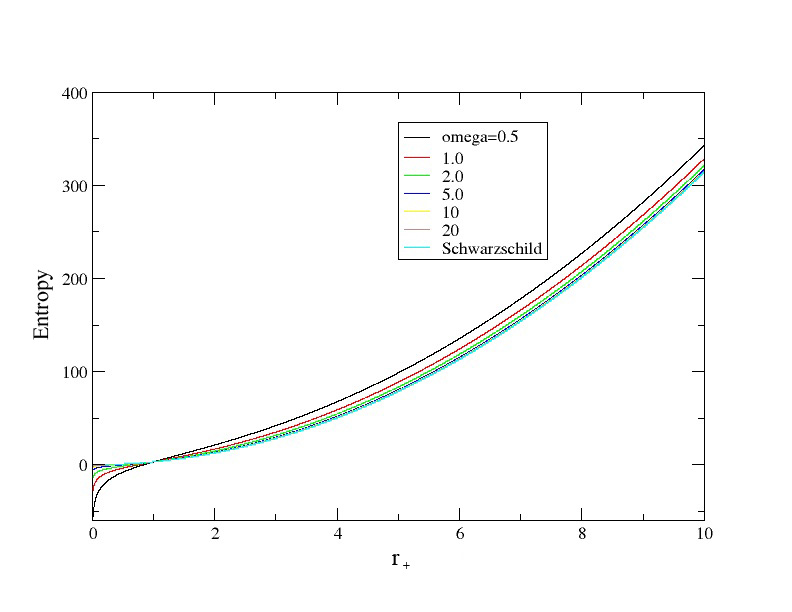}
\par\end{centering}

\caption{\label{fig:Entropy-1}Entropy as a function of the radius of outer
horizon for different values of $\omega$. }

\end{figure}

\begin{figure}[H]
\begin{centering}
\includegraphics[scale=0.4]{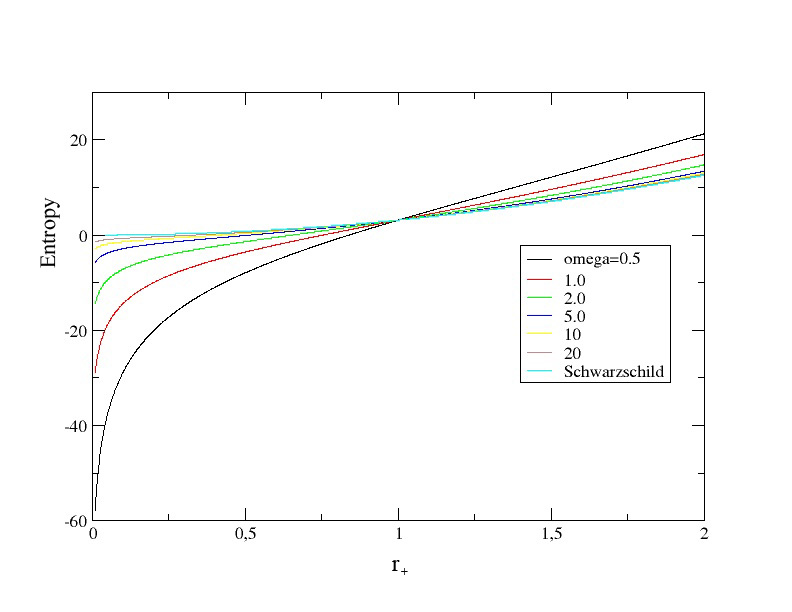}
\par\end{centering}

\caption{\label{fig:Entropy-1-1}Zoom of Figure \ref{fig:Entropy-1}, Showing
the entropy as a function of the radius of outer horizon for different
values of $\omega$. }

\end{figure}

On the other hand, Figure \ref{fig:Entropy-1} shows the behavior
of the entropy as a function of the horizon radius for different values
of $\alpha$ for the Horava-Lifshitz black hole and for Schwarzschild's
solution. When decreasing the value of $\alpha$, i.e. increasing
the value of $\omega$, the entropy curve for Horava-Lifshitz black
hole approaches Schwarzschild's entropy for small $r_{+}$. This behavior
is better seen in Figure \ref{fig:Entropy-1-1}.

\section{Heat Capacity and Hawking-Page Transition}

The heat capacity can be calculated as

\begin{equation}
C=\frac{dM}{dT}=\frac{dM}{dr_{+}}\frac{dr_{+}}{dT}.\end{equation}
Using the temperature (\ref{eq:temperature}) and the mass function
(\ref{eq:massformula}) we have

\begin{equation}
C(r_{+})=-2\pi\frac{(r_{+}^{2}+2\alpha)^{2}(r_{+}^{2}-\alpha)}{r_{+}^{4}-5\alpha r_{+}^{2}-2\alpha^{2}}.\label{eq:heat}\end{equation}

As is noted by Myung \cite{thermodeformedbh} an isolated black hole
like Schwarzschild black hole is never in thermal equilibrium because
it decays by the Hawking radiation. This can be seen from the negative
value of its heat capacity by doing $\alpha=0$ in (\ref{eq:heat}),
$C_{S}=-2\pi r_{+}^{2}$. 

\begin{figure}[H]
\begin{centering}
\includegraphics[scale=0.4]{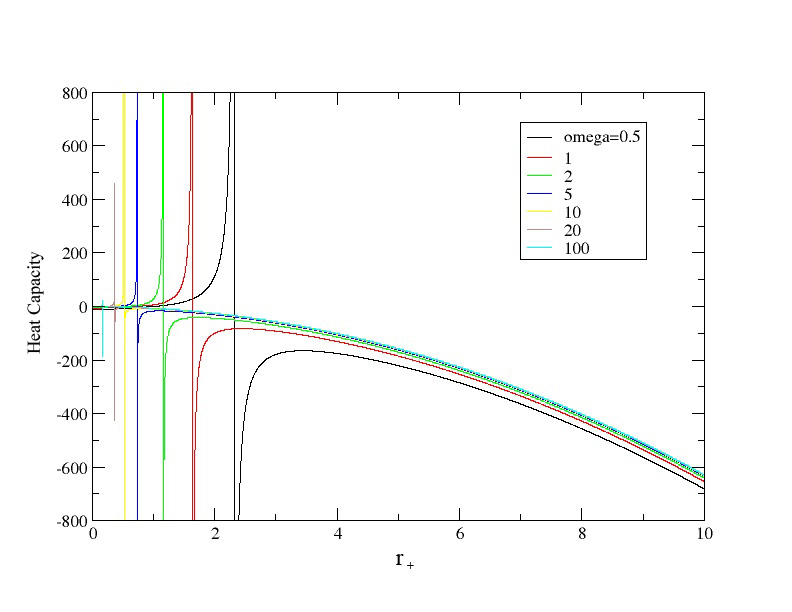}
\par\end{centering}

\caption{\label{fig:heatc}Heat capacity as a function of the radius of outer
horizon for different values of $\omega$. }

\end{figure}

In the case of the Horava-Lifshitz black hole, expression (\ref{eq:heat})
shows that the heat capacity can be negative but also positive, depending
on the value of the parameter $\alpha$. In Figure \ref{fig:heatc}
is easily seen that the heat capacity have positive values for different
values of $\alpha$. The value $r_{+}=r_{m}$ at which the heat capacity
blows is given by

\begin{equation}
r_{m}==\sqrt{\frac{5}{2}\sqrt{33}}\sqrt{\alpha}.\end{equation}
Black holes with $r_{+}<r_{m}$ are local thermodynamically stable
while those with $r_{+}>r_{m}$ are unstable. 

Finally, another interesting question is whether there exists the
Hawking-Page phase transition associated with the Horava-Lifshitz
black hole. In order to discuss the Hawking-Page transition, we have
to calculate the Euclidean action or free energy for the black hole.
The Euclidean action is related with the free energy by

\begin{equation}
I=\frac{1}{T}F,\end{equation}
where $T$ is the temperature of the black hole and the free energy
$F$ is given by

\begin{equation}
F=M-TS.\end{equation}
Using equations (\ref{eq:temperature}), (\ref{eq:massformula}) and
(\ref{eq:entropy}), we find

\begin{equation}
F=\frac{r_{+}^{4}+7\alpha r_{+}^{2}+4\alpha^{2}+4\alpha^{2}\ln\left(r_{+}\right)-4r_{+}^{2}\alpha\ln\left(r_{+}\right)}{4r_{+}(r_{+}^{2}+2\alpha)}.\end{equation}

Note that the free energy is negative only for small enough horizon
radius, which means that large black holes in Horava-Lifshitz gravity
are thermodynamically unstable globally.

\section{Conclusion}

We studied the entropy of black holes in the deformed Horava-Lifshitz
gravity with coupling constant $\lyxmathsym{\textgreek{l}}$. It has
been shown that in the case $\lyxmathsym{\textgreek{l}}=1$, the black
hole resembles the Reissner-Norstrom black hole when it is noted that
the geometric parameter $\alpha=\frac{1}{2\omega}$ in the horizon
radius assumes a similar role as that the electric charge. The entropy
of the Horava-Lifshitz black hole is calculated by assuming that the
first law of thermodynamics is valid for this geometry. The obtained
expression reduces to Schwarzschild's entropy in the limit $\alpha=0$
but differs for other values. Finally we studied the heat capacity
and Hawking-Page phase transition, to show that Black holes with $r_{+}<r_{m}$
are globally thermodynamically stable, while large black holes are
thermodynamically unstable globally.

\end{document}